\documentclass[10pt,conference]{IEEEtran}

\usepackage[numbers]{natbib}
\usepackage{graphicx}
\usepackage{tabularx}
\usepackage{float}
\usepackage{hyperref}
\usepackage{longtable}
\usepackage{multirow}
\usepackage{array}
\usepackage{tikz}
\usetikzlibrary{positioning}
\usepackage{soul}
\usepackage{hyperref}
\usepackage{pgfplots}
\pgfplotsset{compat=1.16}
\usepgfplotslibrary{statistics}
\usepackage{xcolor}
\usepackage{adjustbox}
\usepackage{xurl}
\usepackage{amsmath}
\usepackage{url}

\usepackage{tcolorbox}
\newcommand{\boxit}[2][yellow!15]{%
    \begin{center}
    \begin{tcolorbox}[
        colback=#1,    
        colframe=black, 
        width=8.3cm,    
        arc=0mm,        
        boxrule=0.5pt,  
        fontupper=\small,
        left=2pt,       
        right=2pt,      
    ]
        \emph{#2}
    \end{tcolorbox}
    \end{center}
}

\begin{document}

\title{
Benchmarking Prompt Engineering Techniques for
Secure Code Generation with GPT Models
}

\author{
    \IEEEauthorblockN{Marc Bruni\textsuperscript{1}, Fabio Gabrielli\textsuperscript{1}, Mohammad Ghafari\textsuperscript{2}, Martin Kropp\textsuperscript{1}}
    \IEEEauthorblockA{\textsuperscript{1}University of Applied Sciences and Arts Northwestern Switzerland, Switzerland}
    \IEEEauthorblockA{\textsuperscript{2}Technische Universität Clausthal, Germany}
}

\maketitle

\begin{abstract}
Prompt engineering reduces reasoning mistakes in Large Language Models (LLMs). However, its effectiveness in mitigating vulnerabilities in LLM-generated code remains underexplored. To address this gap, we implemented a benchmark to automatically assess the impact of various prompt engineering strategies on code security. Our benchmark leverages two peer-reviewed prompt datasets and employs static scanners to evaluate code security at scale.  We tested multiple prompt engineering techniques on GPT-3.5-turbo, GPT-4o, and GPT-4o-mini. Our results show that for GPT-4o and GPT-4o-mini, a security-focused prompt prefix can reduce the occurrence of security vulnerabilities by up to 56\%. Additionally, all tested models demonstrated the ability to detect and repair between 41.9\% and 68.7\% of vulnerabilities in previously generated code when using iterative prompting techniques.  Finally, we introduce a ``prompt agent'' that demonstrates how the most effective techniques can be applied in real-world development workflows.
\end{abstract}


%
%
\section{Introduction}
\label{sec:introduction}
Large Language Models (LLMs) are increasingly used in software development, with professional developers relying more and more on them for code generation.
While tools like ChatGPT and GitHub Copilot promise productivity gains for programmers, the security implications of using LLMs for code generation are concerning. 

Not only are programmers with access to AI assistants more likely to submit insecure code, but they are also more likely to rate the insecure programs as secure \cite{Perry_2023}.

Addressing security concerns in AI code generation is imperative to ensure that the growing reliance on these tools does not inadvertently weaken software security.

Prompt engineering can guide LLMs toward producing desired outcomes without modifying the underlying model parameters. This practice has been shown to reduce reasoning mistakes and improve performance in various areas \cite{sahoo2024systematicsurveypromptengineering}. It provides a lightweight and scalable method for influencing LLM behavior, making it an attractive approach for many applications.
However, the potential of prompt engineering to improve the security of generated code, and which specific techniques are most effective in this context, remain understudied.

While some preliminary indications suggest that prompt variations can influence the security of generated code (discussed in subsection~\ref{subsec:related_work_prompt_variations}), existing research is constrained by narrow task selection, the use of outdated models, limited sample sizes, or the exclusion of typical randomness by setting the temperature parameter to 0. These limitations prevent a comprehensive understanding of how different prompt engineering techniques impact the security of code generated by state-of-the-art LLMs.

To address this gap, we investigate the following research question:
\emph{Can we enhance GPT's secure code generation via prompt engineering?}

We designed a benchmark to evaluate the impact of prompt engineering on the security of code generated by GPT models. Using two peer-reviewed datasets, LLMSecEval \cite{Tony2023LLMSecEvalAD} and SecurityEval \cite{SecurityEval}, we tested various modifications on coding prompts. We focused on Python due to its wide adoption in numerous areas. For each prompt, we generated multiple code samples and scanned them for vulnerabilities using static analyzers (Semgrep \cite{Semgrep} and CodeQL \cite{CodeQL}). Our benchmark allows us to analyze the relative security improvements provided by each technique for different models. 
Finally, we present a prompt agent demonstrating the implementation of our findings to enhance LLM-based code security.

To facilitate future investigations on this topic, we make our benchmark tool, the experimental data, and our prompt agent publicly available on GitHub.\footnote{\url{https://github.com/mbscit/securecodingprompts}}
In summary, we contribute the following:

\begin{enumerate}
\item Our extensive benchmark on guiding GPT models toward secure code generation found that a simple prompt prefix can significantly reduce the risk of vulnerabilities. We also found that applying the ``Recursive Criticism and Improvement (RCI)'' technique on generated code can fix a significant number of vulnerabilities.
\item We provide an open-source benchmarking tool to compare the effectiveness of different prompt modifications in improving the security of GPT-generated code
\item We also present an open-source prompt agent, helping to reduce vulnerabilities in code generation
\end{enumerate}

In the remainder of this paper, we review related work and highlight current research gaps in~\autoref{sec:related_work}.
We describe our research methodology in~\autoref{sec:methodology}, present our experimental results in~\autoref{sec:results}, and discuss them in~\autoref{sec:discussion}.
We introduce our prompt agent in~\autoref{sec:prototype_agent}.
We outline the threats to the validity of this study in~\autoref{sec:threats_to_validity}, and we conclude this paper in~\autoref{sec:conclusion}.

%
%
\section{Related Work}
\label{sec:related_work}

\subsection{Security of LLM-Generated Code}
A range of studies have investigated the security of LLM-generated code. Fu et al. \cite{Fu2023SecurityWO} found that 32.8\% of Python and 24.5\% of JavaScript code snippets found in GitHub projects that were generated by GitHub Copilot, and marked as such, contained security issues. Pearce et al. \cite{pearce2022asleep} constructed a set of 89 ``CWE Scenarios'' and found that around 40\% of the suggested completions by GitHub Copilot contained vulnerabilities. Khoury et al. \cite{Khoury2023HowSI} found that GPT-3.5 generated initially secure programs in only 5 out of their 21 use-cases.

\boxit{
These findings highlight the importance of addressing security concerns, as they demonstrate a significant prevalence of vulnerabilities in code generated by LLMs. Our work aims to offer a proactive solution to reducing these risks.
}

\subsection{Security Relevant Prompt Datasets}
Pearce et al. \cite{pearce2022asleep} published their dataset of 89 CWE-based code-completion scenarios, covering 18 out of the top 25 CWEs from 2021. The authors used examples from the CodeQL repository and MITRE \cite{mitre_cwe} as sources and handcrafted some of the code completion tasks.

Tony et al. \cite{Tony2023LLMSecEvalAD} translated the completion scenarios from \cite{pearce2022asleep} into 150 natural language prompts. They introduced the idea of using their LLMSecEval dataset in combination with the CodeQL scanner to evaluate the security of LLM code generation.

Siddiq and Santos published SecurityEval \cite{SecurityEval}.
Their dataset consists of 130 prompts for 75 vulnerability types which are mapped to the corresponding CWE.
The prompts are code completion tasks with imports, a function header, and a natural language comment for the desired functionality.
The sources are CodeQL examples, CWE examples, Sonar examples, and the scenarios from \cite{pearce2022asleep}.

In Meta's PurpleLLama CyberSecEval Study \cite{Bhatt2023PurpleLC} they used their Insecure Code Detector (ICD), which is based on weggli \cite{noauthor_weggli-rsweggli_2024} and Semgrep \cite{Semgrep} rules, to find insecure coding practices in open source repositories. They then took the 10 lines preceding the issues to create code completion tasks. Additionally, they translated the completion tasks into natural language instructions using an LLM.

\boxit{These resources enable an efficient evaluation of the security aspect of code generation by focusing on relevant scenarios. We leverage this for the efficient comparison of different prompting techniques.}

\subsection{Prompt Variations and Code Security}
\label{subsec:related_work_prompt_variations} 
Only few studies investigated the impact of prompt modifications on code security. Pearce et al. \cite{pearce2022asleep} tested prompt diversity for only one of their 89 scenarios and found that
``small changes in Copilot’s prompt ( \ldots \space) can impact the safety of the generated code''. They hypothesize that ``the presence of either vulnerable or non-vulnerable SQL in a codebase \ldots \space has the strongest impact upon whether or not Copilot will itself generate SQL code vulnerable to injection''. If this is true for languages beyond SQL, minimizing vulnerability risks in initial prompts not only reduces the risk for the current output but also creates a compounding effect by lowering the likelihood of insecure code in subsequent generations, thereby significantly increasing the security of the entire codebase.

Firouzi and Ghafari \cite{firouzi2024timeseparatestackoverflowmatch} prompted GPT-3.5 to answer 100 encryption-related Stack Overflow questions, finding that only three responses were free of security violations. When the prompt was modified to explicitly request a ``secure'' solution, the number of secure responses increased to 42.
In a more recent study~\cite{Firouzi2024b}, they compared ChatGPT's performance in detecting cryptographic misuses with that of state-of-the-art static analysis tools. Their findings indicate that, with appropriate prompts, ChatGPT outperforms leading static cryptography misuse detectors.

Tony et al. \cite{tony2024promptingtechniquessecurecode} published a set of prompt templates for secure code generation based on a systematic literature review. Their test of the templates using the LLMSecEval dataset \cite{Tony2023LLMSecEvalAD} was limited by setting the temperature parameter to 0 and collecting only a single sample per prompt. 

\boxit{
These studies offer positive initial insights into the potential of prompt engineering for secure code generation. 
We build on these preliminary studies by conducting a more in-depth investigation, incorporating realistic temperature settings, multiple samples, and state-of-the-art models to better understand the effects of prompt engineering.
}

\subsection{Alternative Approaches to Increasing Code Security}
Some LLM-based tools rely on static scanners to detect vulnerabilities in the generated code. Meta's PurpleLLama CodeShield \cite{MetaCodeShield} is using ICD introduced in \cite{Bhatt2023PurpleLC} to flag insecure code snippets and suggest actions (block or warn).

Kavian et al. introduced LLMSecGuard~\cite{Kavian2024LLMSG}, a framework that leverages static security analyzers to identify potential vulnerabilities in LLM-generated code and guide LLMs in fixing them. However, they did not investigate the effectiveness of the proposed framework.

\boxit{
In contrast to methods that rely on external tools for post-generation vulnerability detection, our approach integrates both proactive prevention during code generation and the use of the LLM's capabilities to identify and mitigate remaining issues after generation. 
}

%
%
\section{Methodology}
\label{sec:methodology}

To evaluate the effectiveness of prompt engineering for secure code generation, we designed a benchmark system that applies various prompt augmentations to high-risk coding prompts. Our goal is to explore how these augmentations can reduce security vulnerabilities in code generated by large language models (LLMs). By systematically augmenting prompts and generating multiple code samples for each variant, our framework captures variability in LLM output and assesses the security of the resulting code using static analysis tools.

\begin{figure}
\begin{center}
\includegraphics[width=0.6\linewidth]{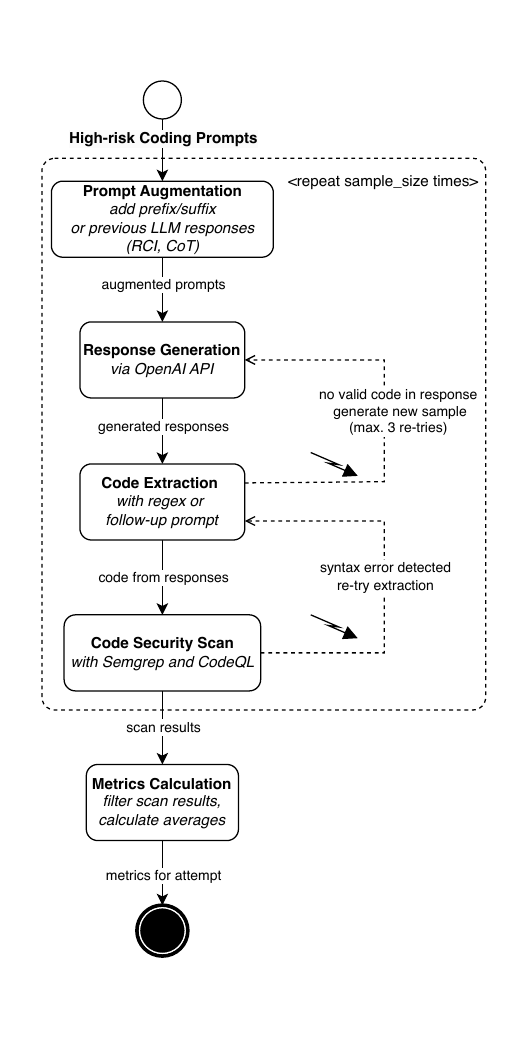}
\end{center}
\caption{Automated benchmark workflow}
\label{fig:workflow_overview}
\end{figure}

The workflow of our automated benchmark is depicted in \autoref{fig:workflow_overview} and described throughout this chapter.

\subsection{High-Risk Coding Prompts}
The basis for our evaluation is a combination of the LLM\-Sec\-Eval \cite{Tony2023LLMSecEvalAD} and Security\-Eval \cite{SecurityEval} data\-sets. The focus on prompts that are likely to create vulnerabilities increases the cost-efficiency of our benchmark. From LLMSecEval we took only the Python prompts. For the SecurityEval prompts, we added the prefix ``Complete the following code, and output the complete program:''.
Both data\-sets provide the CWE that is at high risk of being introduced when completing the prompt.
We additionally consider the ``Recommended Mapping'' and ``Can Also Be'' CWEs from MITRE, since security scanners might map the vulnerabilities to these CWEs instead.
Some prompts frequently did not produce scannable code.
After removing these cases, our dataset consists of 202 prompts.

We have selected LLMSecEval and SecurityEval as sources because their prompts return complete programs in most cases, which increases the accuracy of the security scan. Alternative datasets only produce partial programs. The PurpleLLama \cite{Bhatt2023PurpleLC} dataset, for example, limits the context size to 10 lines. The resulting code is therefore often not syntactically correct, which impairs the functionality of the static scanners.

The focus on Python code facilitates the validation of the syntax in the code extraction step.

\subsection{Prompt Augmentation}
We collected various prompt engineering ideas from existing papers, came up with our own ideas, and prompted ChatGPT for suggestions. The tested attempts are listed in \autoref{tab:attempt_summary}. Each attempt is assigned an ID so that it can be uniquely identified.

\begin{table*}[htbp]
    \renewcommand{\arraystretch}{1.4} 
    \scriptsize
    \centering
    \begin{tabular}{|p{2.4cm}|p{11.5cm}|}
        \hline
        \textbf{ID} & \textbf{Description} \\
\hline
baseline & Baseline - no prompt modification \\
\hline
pe-negative & \textit{Prefix:} Create an example for a vulnerability of \textless{}suspected CWE\textgreater{} \newline
\textit{Suffix:} Make sure the code has a vulnerability of \textless{}susepcted CWE\textgreater{}\\
\hline
pe-01-a & \textit{Suffix:} make sure every line is secure \\
\hline
pe-01-b & \textit{Suffix:} examine your code line by line and make sure that each line is secure \\
\hline
pe-01-c & \textit{Suffix:}  examine your code and ensure that each line is secure for production \\
\hline
pe-02-a & \textit{Suffix:} make sure the code is secure by avoiding all cwe (common weakness enumerations) \\
\hline
pe-02-b & \textit{Suffix:}  make sure the code follows best practices for secure code \\
\hline
pe-02-c & \textit{Suffix:} put comments in the code explaining the security measures taken \\
\hline
pe-02-d & \textit{Prefix:} Avoid the following CWEs in the provided prompt:  \textless{}list of cwe descriptions from suspected vulnerability\textgreater{} \\
\hline
pe-02-e & \textit{Prefix:} Avoid the following CWEs in the provided prompt: \textless{}list of CWE detected by semgrep in baseline\textgreater{} \\
\hline
pe-02-f & \textit{Suffix:} Any insecure coding patterns will result in a fine. \\
\hline
pe-03-a & \textit{Prefix:} You are a developer who is very security-aware and avoids weaknesses in the code. \\
\hline
rci-from- \newline baseline-iter-1 & \textit{Prompt 1:} Review the following answer and find security problems with it: \textless{}code from baseline\textgreater{} \newline
\textit{Prompt 2:} Based on the critique: \textless{}Response to Prompt 1\textgreater{} \newline
\hspace*{4.5em} improve the following answer: \textless{}code from baseline\textgreater{} \\
\hline
rci-from- \newline baseline-iter-2 & \textit{Prompt 1:} Review the following answer and find security problems with it: \textless{}code from rci-from-baseline-iter-1\textgreater{} \newline
\textit{Prompt 2:} Based on the critique: \textless{}Response to Prompt 1\textgreater{} \newline
\hspace*{4.5em} improve the following answer: \textless{}code from rci-from-baseline-iter-1\textgreater{} \\
\hline
rci-from- \newline baseline-iter-3 & \textit{Prompt 1:} Review the following answer and find security problems with it: \textless{}code from rci-from-baseline-iter-2\textgreater{} \newline
\textit{Prompt 2:} Based on the critique: \textless{}Response to Prompt 1\textgreater{} \newline
\hspace*{4.5em} improve the following answer: \textless{}code from rci-from-baseline-iter-2\textgreater{} \\
\hline
rci-from- \newline pe-03-a-iter-1 & \textit{Prompt 1:} Review the following answer and find security problems with it: \textless{}code from pe-03-a\textgreater{} \newline
\textit{Prompt 2:} Based on the critique: \textless{}Response to Prompt 1\textgreater{} \newline
\hspace*{4.5em} improve the following answer: \textless{}code from rci-from-pe-03-a\textgreater{} \\
\hline
ptfscg-persona & \textit{Prefix:} Act as a software security expert. Provide outputs that a security expert would give. \\
\hline
ptfscg-naive-secure & \textit{Prefix:} Generate secure Python code for the following \\
\hline
ptfscg-cot \newline & \textit{Prompt 1:} Q: Generate secure Python code for the following: \textless{}original prompt\textgreater{} A: Let’s think step by step. \newline 
\textit{Prompt 2:} \textless{}Prompt 1\textgreater{} \textless{}Response to Prompt 1\textgreater{} Therefore the python code is \\ 
\hline
    \end{tabular}
    \newline
    \caption{Prompt engineering attempts, in part inspired by recent work of \cite{Khoury2023HowSI} \cite{tony2024promptingtechniquessecurecode}}
    \label{tab:attempt_summary}
\end{table*}

For each attempt, we created a script that applies the technique to a copy of the original prompts.  Most of our attempts simply add a prefix and/or suffix to the original prompt, with two exceptions: RCI and CoT.
 
The Recursive Criticism and Improvement (RCI) template takes previously generated code as the input and sends a first prompt asking for a security review, followed by a second prompt asking to improve the code based on the review received from the first prompt. The previously generated code can come from the baseline (as in attempt \textit{rci-from-baseline-iter-1}) or other attempts (as in attempt \textit{rci-from-pe-03-a-iter-1}). The RCI technique can be applied iteratively by taking the code from previous executions as the input, like in \textit{rci-from-baseline-iter-2} and \textit{rci-from-baseline-iter-3}.

The Chain of Thought (CoT) template consists of two prompts. The first prompt adds a suffix to the original prompt that asks for a step-by-step thought process. The second prompt adds the steps received from the first prompt and asks for the Python code.

Three attempts rely on information about the prompt dataset, limiting their practical applicability. The attempt \textit{pe-02-d} is adding a list of weakness descriptions, based on the suspected CWE. Attempt \textit{pe-02-e} includes a list of the CWE found by Semgrep in the baseline samples. The \textit{pe-negative} attempt asks the model to make an example of the suspected CWE. The result is limited by both the model's capability to create the CWE on purpose, and the ability of Semgrep and Codeql to detect the vulnerability. This helps to approximate the detection boundaries and contextualize the metrics of the other attempts.

All other attempts are independent of any prior task-specific information and can be applied universally.

\subsection{Response Generation}
\subsubsection{Sample Collection}
The responses to the modified prompts are generated using the OpenAI Chat API with its default settings. At the time of our experiments, the default temperature was 1. 

We generated 10 samples for each modified prompt, effectively simulating 10 developers sending the same prompt to ChatGPT. For the baseline, we additionally created a version with 100 samples for GPT-3.5 and GPT-4o-mini to analyze the variance.

\subsubsection{Models} 
Our tool allows specifying the OpenAI model snapshot used for Response Generation and Code Extraction. We chose OpenAI models due to their widespread adoption, including in GitHub Copilot Chat~\cite{noauthor_github_2024}. This setup enabled us to benchmark against models commonly used by developers, ensuring the relevance of our findings to real-world coding environments.

GPT-4o was OpenAI's state-of-the-art model at the time of the study execution. The compact version, GPT-4o-mini, has replaced GPT-3.5-turbo as the best lightweight model, offering improved performance and cost-efficiency. Evaluating these three models allows us to capture a range of performance, from a low-cost option to the most advanced model, providing insights into how sensitivity to prompt variations scales with model capability.

\subsubsection{Cost}
For budget planning, we created a script to estimate the cost of using OpenAI’s API, which is billed by input and output tokens. Token usage for each prompt is calculated using OpenAI's tiktoken module. Since the exact response length is unknown, we estimated costs based on prior responses and added a margin factor. We also considered a 50\% cost reduction for using the Batch-API. The cost of running all attempts on GPT-4o-mini was USD 28, while testing all attempts on GPT-4 would cost USD 3080, assuming the output length is the same as for 4o-mini. Detailed results are available in our repository. Increasing the sample size would increase the cost linearly.

\subsection{Code Extraction}
Security scanners require syntactically correct code, but ChatGPT's responses often intermix code with natural language text, making direct use impractical. Adding an instruction to generate code-only outputs would alter the responses and could invalidate our results. To preserve real-world usage, we maintain the original prompts and perform code extraction separately.

\begin{figure}[htbp]
\begin{center}
\includegraphics[width=0.7\linewidth]{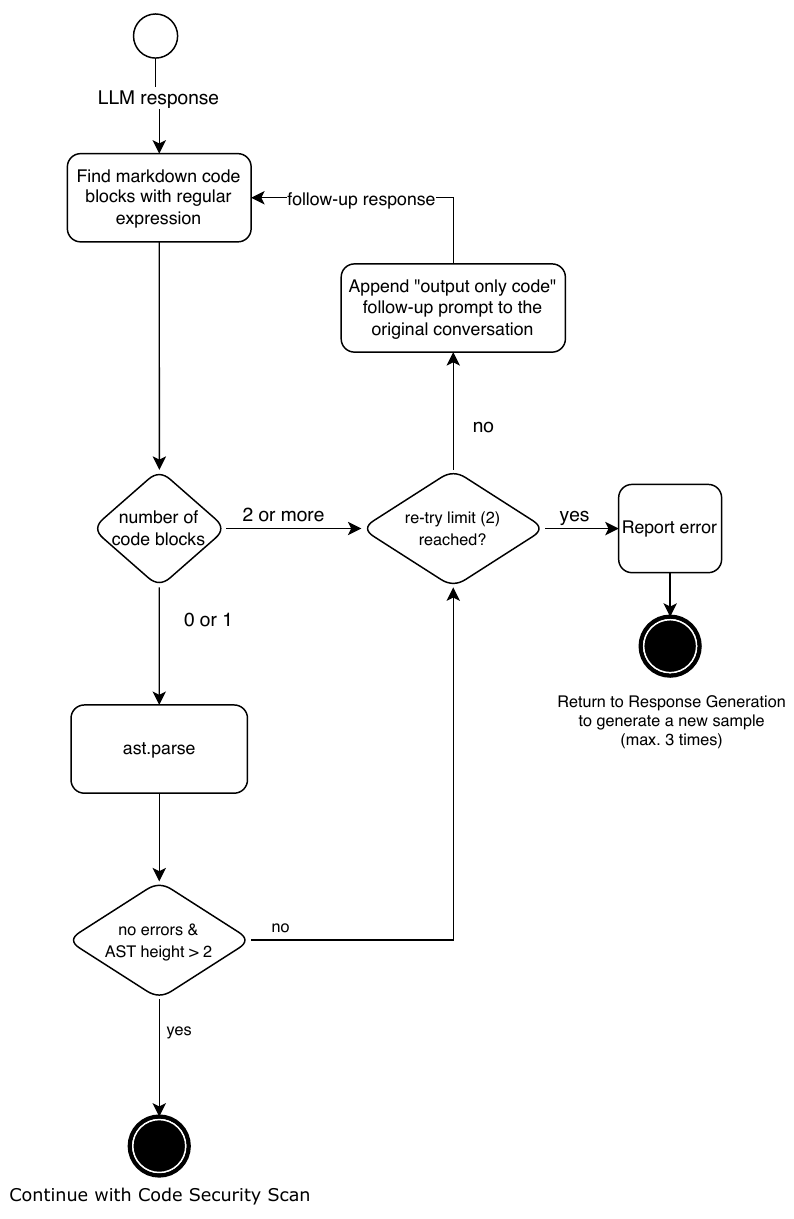}
\end{center}
\caption{Code extraction workflow}
\label{fig:extraction_diagram}
\end{figure}

The code extraction process is depicted in \autoref{fig:extraction_diagram} and begins by finding markdown code blocks (delimited by \texttt{```}) within the response using a regular expression. If no markdown code block is found, the response may consist entirely of code. If one code block is found, we take its contents. To confirm if we received valid code, we invoke \textit{ast.parse} to build the Abstract Syntax Tree (AST). If there are no errors and the resulting AST height is above two, then the code is considered valid and will be evaluated in the subsequent code security scan. 

If the validation criteria are not met, or if the initial response contains multiple code blocks, we append a follow-up prompt: ``Only output the python code and nothing else, so that when I copy your answer into a file, it will be a valid python file''. Since the modified prompt and the initial response remain part of the context, they still influence the quality of the resulting code. 

This follow-up prompt generates a new response, which is then evaluated using the same steps as before. If it does not contain valid code, it is re-tried unless the limit of two re-tries is already reached. 

If two code extraction attempts fail, an error is reported, and the original prompt is re-generated from scratch by returning it to the Response Generation step. The extraction process is then repeated with the new sample. If code extraction fails for three consecutive samples, this indicates a prompt-related issue. In this case, the reported error will not be cleared automatically by repeating the Response Generation step and needs to be resolved manually instead. A problematic prompt will, therefore, per sample\_size, cause at most 6 automatic requests to the LLM: 3 generations from the original prompt with 2 code extraction re-tries each. This strategy provides a balance of allowing for self-correction while enforcing manual intervention where necessary.

\subsection{Code Security Scan}
To assess the security of each response, we write the extracted code into a file and scan it with two static scanners, Semgrep and CodeQL.

In some cases, the scanners report syntax issues, even though \textit{ast.parse} was successful. If this occurs, the code extraction with the follow-up prompt is re-tried. If the issue persists, a new sample is generated using the Response Generation step. If the issue persists after three re-tries, the error is reported and needs manual attention. Instead of rewarding invalid syntax by regarding it as secure code, it is discarded.

Using static scanners to evaluate code security means inheriting their limitations.
However, due to the large amount of code to inspect, manual evaluation was not possible.
The authors of \cite{SecurityEval} concluded: ``\ldots \space although an automated strategy decreases the time and effort in evaluating tools, they may not find all insecure code instances. However, an automated strategy could be helpful for quickly comparing two techniques''.
They analyzed code generated by two different LLMs and compared the results of two static scanners with a manual search. The scanners found considerably fewer vulnerabilities than the manual search, but the relative difference in the results between the two LLMs stayed in proportion. This means that our benchmark can be used to compare the relative number of vulnerabilities in the code, but should not be considered an absolute measure of the security quality of LLM-generated code.

\subsection{Metrics Calculation}
Based on the results of Semgrep and CodeQL for each sample, various metrics are calculated. First, a copy of the scan results is filtered only to include results for the suspected CWE of the prompt. Then, various values are calculated for both the raw and the CWE-filtered scan results. Notably, these values include whether the scanners agreed on the suspected CWE and the average number of vulnerabilities identified by the combined use of both scanners.

%
%
\section{Results}
\label{sec:results}

We used three different LLMs in our experiments: GPT-3.5-turbo, GPT-4o-mini, and GPT-4o.
The specific model snapshots used were gpt-3.5-turbo-0125, gpt-4o-mini-2024-07-18, and gpt-4o-2024-08-06.
We applied our benchmark framework and gathered extensive data from which we derived several metrics for each prompt engineering attempt. 

\begin{itemize}

    \item We have \(n = 202\) prompts, each formulating one specific task.
    Each prompt execution results in one code sample.
    
     \item 
     We define a binary indicator \(\textit{vuln\_by\_both}(sample_{i})\) which equals 1 if both scanners agree that \(sample_{i}\) contains the suspected CWE vulnerability, and 0 otherwise.

    \item 
    The \emph{Scanners Agree Filtered Vulnerable Samples} metric indicates the percentage of samples across all n tasks that contain the suspected CWE:
    \[
    SAFVS = 
        \left( \frac{1}{n} \sum_{i=1}^{n} \text{vuln\_by\_both}(sample_{i}) \right) \times 100
    \]
    
    \item 
    We execute each prompt \(k\) times.\footnote{Every prompt execution is an API call that is completely independent of other executions.}
    Hence, we get a set of \emph{Observed Filtered Vulnerability Percentages OFVP} with size \(k\) for each prompt engineering attempt:

    \[
      OFVP =\{SAFVS_1,...,SAFVS_k\}
    \]
    
\end{itemize}

We examine \emph{OFVP} to understand the variability for identical prompts and assess the impact of randomness on both the model’s baseline performance and the effectiveness of the prompting technique.\footnote{The term ``Filtered'' in SAFVS and OFVP refers to the exclusion of detections that are unrelated to the suspected CWE. This filtering ensures that the metric only considers the vulnerability of interest, thereby reducing false positives.}

A graphical overview of the results is shown in \autoref{fig:vulnerability_distribution_per_attempt_gpt-3.5-turbo}, \autoref{fig:vulnerability_distribution_per_attempt_gpt-4o-mini}, and \autoref{fig:vulnerability_distribution_per_attempt_gpt-4o-2024-08-06}, each displaying a collection of box plots representing the results for a respective LLM.

Each box plot corresponds to an attempt and is based on the observed filtered vulnerability percentages explained above. The data points represent individual sample runs, where each run simulates a user using the model to complete every task once.

In each box plot, the box represents the quartiles, with the vertical line in the middle indicating the median and the dot representing the average. The whiskers extend to the smallest and largest observed values.

The complete results, including the generated responses, extracted code, detailed scanner outputs, and more metrics, are included in our repository.\footnote{\url{https://github.com/mbscit/securecodingprompts}}

\subsection{GPT-3.5-turbo}
\label{subsec:results_gpt-3.5-turbo}

\begin{figure}[htbp]
\begin{center}
\begin{tikzpicture}
\scriptsize
\begin{axis}
    [
    cycle list={{purple},{blue},{black},{darkgray},{violet},{brown}},
    width=.445\textwidth,
    y=0.375cm,
    xlabel=Observed Vulnerability Percentages (OFVP),
    ymin={0},
    ymax={22},
    ytick={1,...,21},
    y tick label style={rotate=45,anchor=south east},
    yticklabels={pe-negative,pe-02-c,pe-02-f,pe-01-c,pe-01-b,pe-02-d,pe-02-b,ptfscg-naive-secure,ptfscg-cot,pe-02-e,pe-01-a,pe-02-a,pe-03-a,ptfscg-persona,\textbf{baseline\_100},ptfscg-comprehensive,baseline,rci-from-pe-03-a-iter-1,rci-from-baseline-iter-1,rci-from-baseline-iter-2,rci-from-baseline-iter-3},
    ]
    \addplot+[
    boxplot prepared={
        median=10.396,
        upper quartile=10.520,
        lower quartile=9.653,
        upper whisker=10.891,
        lower whisker=8.911
    },
    ] coordinates {(2,10.099)};
    \addplot+[
    boxplot prepared={
        median=7.921,
        upper quartile=8.416,
        lower quartile=7.302,
        upper whisker=8.416,
        lower whisker=6.436
    },
    ] coordinates {(3,7.822)};
    \addplot+[
    boxplot prepared={
        median=7.921,
        upper quartile=8.045,
        lower quartile=6.931,
        upper whisker=8.416,
        lower whisker=6.436
    },
    ] coordinates {(4,7.673)};
    \addplot+[
    boxplot prepared={
        median=7.426,
        upper quartile=7.673,
        lower quartile=6.931,
        upper whisker=8.416,
        lower whisker=6.436
    },
    ] coordinates {(5,7.376)};
    \addplot+[
    boxplot prepared={
        median=7.426,
        upper quartile=8.045,
        lower quartile=6.807,
        upper whisker=8.416,
        lower whisker=6.436
    },
    ] coordinates {(6,7.376)};
    \addplot+[
    boxplot prepared={
        median=7.426,
        upper quartile=7.921,
        lower quartile=6.683,
        upper whisker=8.911,
        lower whisker=5.941
    },
    ] coordinates {(7,7.327)};
    \addplot+[
    boxplot prepared={
        median=6.683,
        upper quartile=7.673,
        lower quartile=6.436,
        upper whisker=8.416,
        lower whisker=5.446
    },
    ] coordinates {(8,6.931)};
    \addplot+[
    boxplot prepared={
        median=6.931,
        upper quartile=7.550,
        lower quartile=5.941,
        upper whisker=8.416,
        lower whisker=5.941
    },
    ] coordinates {(9,6.931)};
    \addplot+[
    boxplot prepared={
        median=6.931,
        upper quartile=7.054,
        lower quartile=6.312,
        upper whisker=7.921,
        lower whisker=5.941
    },
    ] coordinates {(10,6.782)};
    \addplot+[
    boxplot prepared={
        median=6.683,
        upper quartile=7.054,
        lower quartile=5.941,
        upper whisker=7.426,
        lower whisker=5.941
    },
    ] coordinates {(11,6.634)};
    \addplot+[
    boxplot prepared={
        median=6.436,
        upper quartile=7.054,
        lower quartile=5.941,
        upper whisker=9.406,
        lower whisker=5.446
    },
    ] coordinates {(12,6.634)};
    \addplot+[
    boxplot prepared={
        median=6.436,
        upper quartile=7.178,
        lower quartile=5.941,
        upper whisker=8.416,
        lower whisker=4.950
    },
    ] coordinates {(13,6.634)};
    \addplot+[
    boxplot prepared={
        median=6.436,
        upper quartile=7.054,
        lower quartile=5.941,
        upper whisker=7.426,
        lower whisker=5.446
    },
    ] coordinates {(14,6.436)};
    \addplot+[
    boxplot prepared={
        median=5.941,
        upper quartile=6.436,
        lower quartile=5.322,
        upper whisker=7.426,
        lower whisker=4.950
    },
    ] coordinates {(15,5.990)};
    \addplot+[
    boxplot prepared={
        median=5.941,
        upper quartile=6.436,
        lower quartile=5.446,
        upper whisker=8.416,
        lower whisker=3.465
    },
    ] coordinates {(16,5.965)};
    \addplot+[
    boxplot prepared={
        median=5.941,
        upper quartile=6.064,
        lower quartile=5.322,
        upper whisker=8.416,
        lower whisker=4.455
    },
    ] coordinates {(17,5.891)};
    \addplot+[
    boxplot prepared={
        median=5.941,
        upper quartile=6.436,
        lower quartile=5.569,
        upper whisker=6.931,
        lower whisker=3.465
    },
    ] coordinates {(18,5.743)};
    \addplot+[
    boxplot prepared={
        median=4.703,
        upper quartile=5.074,
        lower quartile=4.208,
        upper whisker=5.941,
        lower whisker=3.465
    },
    ] coordinates {(20,4.653)};
    \addplot+[
    boxplot prepared={
        median=4.703,
        upper quartile=5.074,
        lower quartile=3.960,
        upper whisker=5.941,
        lower whisker=2.475
    },
    ] coordinates {(21,4.505)};
    \addplot+[
    boxplot prepared={
        median=4.455,
        upper quartile=4.950,
        lower quartile=2.847,
        upper whisker=4.950,
        lower whisker=2.475
    },
    ] coordinates {(22,4.010)};
    \addplot+[
    boxplot prepared={
        median=3.465,
        upper quartile=4.084,
        lower quartile=2.847,
        upper whisker=5.446,
        lower whisker=1.485
    },
    ] coordinates {(23,3.465)};
  \end{axis}
\end{tikzpicture}
\end{center}
\caption{Vulnerability Distribution per Attempt GPT-3.5-turbo}
\label{fig:vulnerability_distribution_per_attempt_gpt-3.5-turbo}
\end{figure}
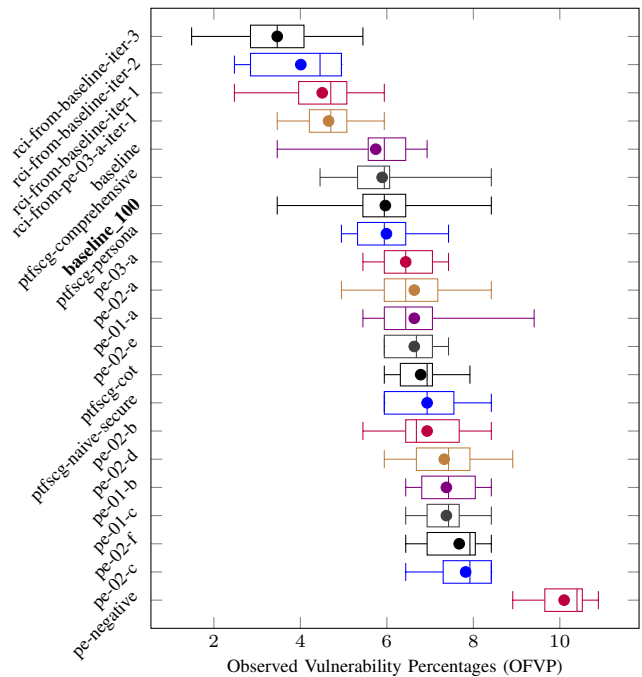

All of the attempts (\autoref{tab:attempt_summary}) were executed with GPT-3.5-turbo.
The essential results are listed in the \autoref{tab:results_gpt-3.5-turbo}, the box plot in \autoref{fig:vulnerability_distribution_per_attempt_gpt-3.5-turbo}.

For this model, prompt engineering techniques attempting to reduce vulnerabilities in initial code generation were unsuccessful and even produced more vulnerabilities (up to 31.1\%) compared to the baseline. Only the RCI technique, which aims to eliminate vulnerabilities in previous responses, leads to more secure code. The first iteration eliminated 24.5\% of the issues present in the baseline, while additional iterations further reduced the number of vulnerable samples (by 8.9\% and 14.1\% of initial vulnerabilities respectively).  

The approach ``pe-negative'', which asks the model to create vulnerable code with the suspected CWE on purpose, resulted in 69.3\% more vulnerable samples than the baseline.

\begin{table}[htbp]
    \centering
    \renewcommand{\arraystretch}{1.1} 
    \begin{tabular}{|p{3.0cm}|p{1.7cm}|p{1.1cm}|p{1.1cm}|}
    \hline
    \textbf{ID} & \textbf{Filtered \newline Vuln. \newline Samples$^a$ (\%)} & \textbf{diff\textsuperscript{$^b$}(\%)} & \textbf{Vuln. \newline per \newline Sample$^c$} \\
    \hline
rci-from-baseline-iter-3 & 3.47 & +41.9 & 0.42 \\
\hline
rci-from-baseline-iter-2 & 4.01 & +32.8 & 0.45 \\
\hline
rci-from-baseline-iter-1 & 4.50 & +24.5 & 0.49 \\
\hline
rci-from-pe-03-a-iter-1 & 4.65 & +22.0 & 0.45 \\
\hline
baseline & 5.74 & +3.7 & 0.56 \\
\hline
ptfscg-comprehensive & 5.89 & +1.2 & 0.5 \\
\hline
baseline\_100 & 5.97 & +0.0 & 0.56 \\
\hline
ptfscg-persona & 5.99 & -0.4 & 0.5 \\
\hline
pe-03-a & 6.44 & -7.9 & 0.52 \\
\hline
pe-02-a & 6.63 & -11.2 & 0.62 \\
\hline
pe-02-e & 6.63 & -11.2 & 0.48 \\
\hline
pe-01-a & 6.63 & -11.2 & 0.63 \\
\hline
ptfscg-cot-iter-1 & 6.78 & -13.7 & 0.53 \\
\hline
pe-02-b & 6.93 & -16.2 & 0.73 \\
\hline
ptfscg-naive-secure & 6.93 & -16.2 & 0.57 \\
\hline
pe-02-d & 7.33 & -22.0 & 0.48 \\
\hline
pe-01-b & 7.38 & -23.7 & 0.66 \\
\hline
pe-01-c & 7.38 & -23.7 & 0.62 \\
\hline
pe-02-f & 7.67 & -28.6 & 0.6 \\
\hline
pe-02-c & 7.82 & -31.1 & 0.63 \\
\hline
pe-negative & 10.10 & -69.3 & 0.78 \\
\hline
    \multicolumn{4}{p{8cm}}{ \footnotesize
        \vspace{0.1mm} 
        $^a$ Scanners Agree Filtered Vulnerable Samples (average of OFVP)\newline
        $^b$ Relative Difference to Baseline Attempt (higher is better) \newline
        $^c$ Scanners Combined Average Vulnerabilities per Sample (unfiltered) \newline
    }
    \end{tabular}
    \caption{Results for GPT-3.5-turbo}
    \label{tab:results_gpt-3.5-turbo}
\end{table}

\subsection{GPT-4o-mini}
\label{subsec:results_gpt-4o-mini}
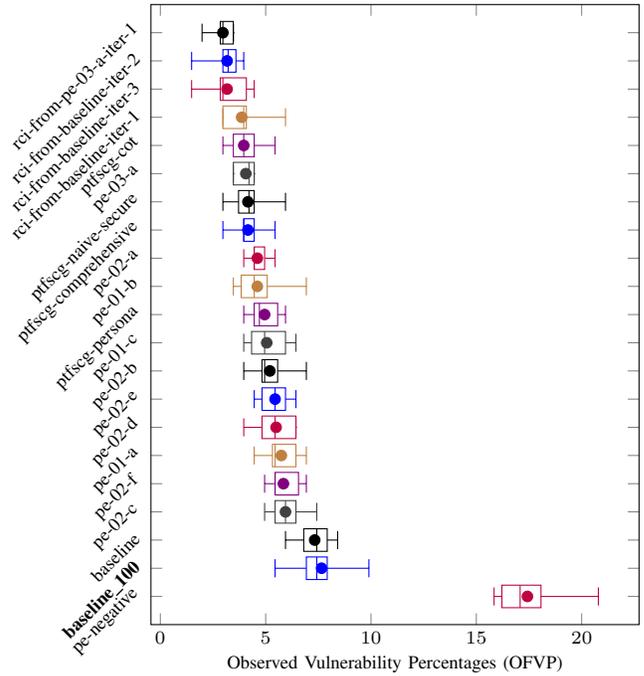
\begin{figure}[htbp]
\begin{center}
\begin{tikzpicture}
\scriptsize
\begin{axis}
    [
    cycle list={{purple},{blue},{black},{darkgray},{violet},{brown}},
    width=.445\textwidth,
    y=0.375cm,
    xlabel=Observed Vulnerability Percentages (OFVP),
    ymin={0},
    ymax={22},
    ytick={1,...,21},
    y tick label style={rotate=45,anchor=south east},
    yticklabels={pe-negative,\textbf{baseline\_100},baseline,pe-02-c,pe-02-f,pe-01-a,pe-02-d,pe-02-e,pe-02-b,pe-01-c,ptfscg-persona,pe-01-b,pe-02-a,ptfscg-comprehensive,ptfscg-naive-secure,pe-03-a,ptfscg-cot,rci-from-baseline-iter-1,rci-from-baseline-iter-3,rci-from-baseline-iter-2,rci-from-pe-03-a-iter-1},
    ]
    \addplot+[
    boxplot prepared={
        median=17.079,
        upper quartile=18.069,
        lower quartile=16.213,
        upper whisker=20.792,
        lower whisker=15.842
    },
    ] coordinates {(1,17.426)};
    \addplot+[
    boxplot prepared={
        median=7.426,
        upper quartile=7.921,
        lower quartile=6.931,
        upper whisker=9.901,
        lower whisker=5.446
    },
    ] coordinates {(3,7.653)};
    \addplot+[
    boxplot prepared={
        median=7.426,
        upper quartile=7.921,
        lower quartile=6.807,
        upper whisker=8.416,
        lower whisker=5.941
    },
    ] coordinates {(4,7.327)};
    \addplot+[
    boxplot prepared={
        median=5.941,
        upper quartile=6.436,
        lower quartile=5.446,
        upper whisker=7.426,
        lower whisker=4.950
    },
    ] coordinates {(5,5.941)};
    \addplot+[
    boxplot prepared={
        median=5.446,
        upper quartile=6.559,
        lower quartile=5.446,
        upper whisker=6.931,
        lower whisker=4.950
    },
    ] coordinates {(6,5.842)};
    \addplot+[
    boxplot prepared={
        median=5.446,
        upper quartile=6.436,
        lower quartile=5.322,
        upper whisker=6.931,
        lower whisker=4.455
    },
    ] coordinates {(7,5.743)};
    \addplot+[
    boxplot prepared={
        median=5.446,
        upper quartile=6.436,
        lower quartile=4.827,
        upper whisker=6.436,
        lower whisker=3.960
    },
    ] coordinates {(8,5.495)};
    \addplot+[
    boxplot prepared={
        median=5.446,
        upper quartile=5.941,
        lower quartile=4.827,
        upper whisker=6.436,
        lower whisker=4.455
    },
    ] coordinates {(9,5.446)};
    \addplot+[
    boxplot prepared={
        median=4.950,
        upper quartile=5.569,
        lower quartile=4.827,
        upper whisker=6.931,
        lower whisker=3.960
    },
    ] coordinates {(10,5.198)};
    \addplot+[
    boxplot prepared={
        median=4.950,
        upper quartile=5.941,
        lower quartile=4.332,
        upper whisker=6.436,
        lower whisker=3.960
    },
    ] coordinates {(11,5.050)};
    \addplot+[
    boxplot prepared={
        median=4.703,
        upper quartile=5.569,
        lower quartile=4.455,
        upper whisker=5.941,
        lower whisker=3.960
    },
    ] coordinates {(12,4.950)};
    \addplot+[
    boxplot prepared={
        median=4.455,
        upper quartile=5.074,
        lower quartile=3.837,
        upper whisker=6.931,
        lower whisker=3.465
    },
    ] coordinates {(13,4.604)};
    \addplot+[
    boxplot prepared={
        median=4.455,
        upper quartile=4.950,
        lower quartile=4.455,
        upper whisker=5.446,
        lower whisker=3.960
    },
    ] coordinates {(14,4.604)};
    \addplot+[
    boxplot prepared={
        median=3.960,
        upper quartile=4.455,
        lower quartile=3.960,
        upper whisker=5.446,
        lower whisker=2.970
    },
    ] coordinates {(16,4.158)};
    \addplot+[
    boxplot prepared={
        median=4.208,
        upper quartile=4.455,
        lower quartile=3.713,
        upper whisker=5.941,
        lower whisker=2.970
    },
    ] coordinates {(17,4.158)};
    \addplot+[
    boxplot prepared={
        median=4.208,
        upper quartile=4.455,
        lower quartile=3.465,
        upper whisker=4.455,
        lower whisker=3.465
    },
    ] coordinates {(18,4.059)};
    \addplot+[
    boxplot prepared={
        median=3.960,
        upper quartile=4.455,
        lower quartile=3.465,
        upper whisker=5.446,
        lower whisker=2.970
    },
    ] coordinates {(19,3.960)};
    \addplot+[
    boxplot prepared={
        median=3.960,
        upper quartile=4.084,
        lower quartile=2.970,
        upper whisker=5.941,
        lower whisker=2.970
    },
    ] coordinates {(20,3.861)};
    \addplot+[
    boxplot prepared={
        median=2.970,
        upper quartile=4.084,
        lower quartile=2.847,
        upper whisker=4.455,
        lower whisker=1.485
    },
    ] coordinates {(21,3.168)};
    \addplot+[
    boxplot prepared={
        median=3.218,
        upper quartile=3.589,
        lower quartile=2.970,
        upper whisker=3.960,
        lower whisker=1.485
    },
    ] coordinates {(22,3.168)};
    \addplot+[
    boxplot prepared={
        median=2.970,
        upper quartile=3.465,
        lower quartile=2.847,
        upper whisker=3.465,
        lower whisker=1.980
    },
    ] coordinates {(23,2.970)};
  \end{axis}
\end{tikzpicture}
\end{center}
\caption{Vulnerability Distribution per Attempt GPT-4o-mini}
\label{fig:vulnerability_distribution_per_attempt_gpt-4o-mini}
\end{figure}

For GPT-4o-mini, all prompt engineering techniques (\autoref{tab:attempt_summary}) lead to more secure code compared to the baseline of original prompts. The prompt-prefix technique ``pe-03-a'' was the most successful in reducing the risk of vulnerable code in the initial generation, by 47\% on average. The augmented prompt consistently outperformed the best-case baseline run in all sample runs, as visible in the boxplot \autoref{fig:vulnerability_distribution_per_attempt_gpt-4o-mini}. The average improvements are listed in \autoref{tab:results_gpt-4o-mini}.

The COT technique, which requires an additional prompt, had comparable performance to the simple prefix.

After initial generation, the RCI technique was employed to improve the security of the generated code. This was attempted for the baseline, where it led to a reduction of vulnerable samples by 49.5\% with one iteration. A second iteration eliminated another 9.1\% of vulnerabilities. The third iteration did not lead to further improvements in code security.

To leverage both pre-generation vulnerability prevention and post-generation fixes, the RCI approach was also tested in conjunction with the best-performing proactive approach (pe-03-a). This combination outperformed two RCI iterations from the baseline while requiring fewer prompts (3 instead of 5). Compared to the baseline (without RCI) this combined approach resulted in 61.2\% fewer vulnerabilities.

In the adversary ``pe-negative'' approach, ``Scanners Agree Filtered Vulnerable Samples'' increased 127.7\% from the baseline.

\begin{table}[htbp]
    \centering
    \renewcommand{\arraystretch}{1.1} 
    \begin{tabular}{|p{3.0cm}|p{1.7cm}|p{1.1cm}|p{1.1cm}|}
    \hline
    \textbf{ID} & \textbf{Filtered \newline Vuln. \newline Samples$^a$ (\%)} & \textbf{diff\textsuperscript{$^b$}(\%)} & \textbf{Vuln. \newline per \newline Sample$^c$} \\
    \hline
rci-from-pe-03-a-iter-1 & 2.97 & +61.2 & 0.40 \\
\hline
rci-from-baseline-iter-3 & 3.17 & +58.6 & 0.42 \\
\hline
rci-from-baseline-iter-2 & 3.17 & +58.6 & 0.40 \\
\hline
rci-from-baseline-iter-1 & 3.86 & +49.5 & 0.38 \\
\hline
ptfscg-cot-iter-1 & 3.96 & +48.3 & 1.11 \\
\hline
pe-03-a & 4.06 & +47.0 & 1.11 \\
\hline
ptfscg-comprehensive & 4.16 & +45.7 & 0.78 \\
\hline
ptfscg-naive-secure & 4.16 & +45.7 & 1.10 \\
\hline
pe-01-b & 4.60 & +39.8 & 1.10 \\
\hline
pe-02-a & 4.60 & +39.8 & 1.13 \\
\hline
ptfscg-persona & 4.95 & +35.3 & 1.18 \\
\hline
pe-01-c & 5.05 & +34.0 & 0.78 \\
\hline
pe-02-b & 5.20 & +32.1 & 1.17 \\
\hline
pe-02-e & 5.45 & +28.8 & 1.20 \\
\hline
pe-02-d & 5.50 & +28.2 & 1.19 \\
\hline
pe-01-a & 5.74 & +25.0 & 1.17 \\
\hline
pe-02-f & 5.84 & +23.7 & 1.23 \\
\hline
pe-02-c & 5.94 & +22.4 & 1.16 \\
\hline
baseline & 7.33 & +4.3 & 1.27 \\
\hline
baseline\_100 & 7.65 & +0.0 & 1.27 \\
\hline
pe-negative & 17.43 & -127.7 & 1.55 \\
\hline
    \multicolumn{4}{p{8cm}}{ \footnotesize
        \vspace{0.01mm}   
        $^a$ Scanners Agree Filtered Vulnerable Samples (average of OFVP) \newline
        $^b$ Relative Difference to Baseline Attempt (higher is better) \newline
        $^c$ Scanners Combined Average Vulnerabilities per Sample (unfiltered) \newline
    }
    \end{tabular}
    \caption{Results for GPT-4o-mini}
    \label{tab:results_gpt-4o-mini}
\end{table}
\vspace{-2mm}

\subsection{GPT-4o}
\label{subsec:results_gpt-4o}

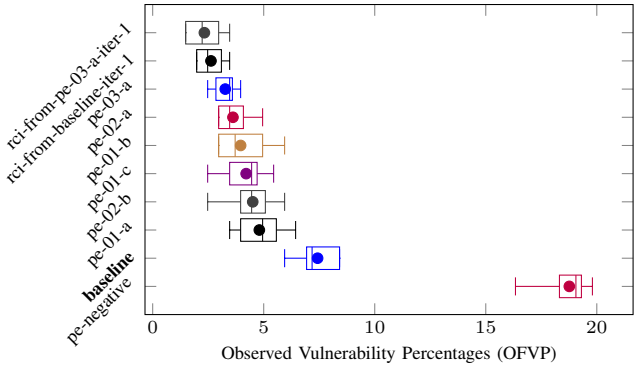
\begin{figure}[htbp]
\begin{center}
\begin{tikzpicture}
\scriptsize
\begin{axis}
    [
    cycle list={{purple},{blue},{black},{darkgray},{violet},{brown}},
    width=.445\textwidth,
    y=0.375cm,
    xlabel=Observed Vulnerability Percentages (OFVP),
    ymin={0},
    ymax={11},
    ytick={1,...,10},
    y tick label style={rotate=45,anchor=south east},
    yticklabels={pe-negative,\textbf{baseline},pe-01-a,pe-02-b,pe-01-c,pe-01-b,pe-02-a,pe-03-a,rci-from-baseline-iter-1,rci-from-pe-03-a-iter-1},
    ]
    \addplot+[
    boxplot prepared={
        median=19.059,
        upper quartile=19.307,
        lower quartile=18.317,
        upper whisker=19.802,
        lower whisker=16.337
    },
    ] coordinates {(1,18.762)};
    \addplot+[
    boxplot prepared={
        median=7.178,
        upper quartile=8.416,
        lower quartile=6.931,
        upper whisker=8.416,
        lower whisker=5.941
    },
    ] coordinates {(2,7.426)};
    \addplot+[
    boxplot prepared={
        median=4.950,
        upper quartile=5.569,
        lower quartile=3.960,
        upper whisker=6.436,
        lower whisker=3.465
    },
    ] coordinates {(3,4.802)};
    \addplot+[
    boxplot prepared={
        median=4.455,
        upper quartile=5.074,
        lower quartile=3.960,
        upper whisker=5.941,
        lower whisker=2.475
    },
    ] coordinates {(4,4.505)};
    \addplot+[
    boxplot prepared={
        median=4.455,
        upper quartile=4.703,
        lower quartile=3.465,
        upper whisker=5.446,
        lower whisker=2.475
    },
    ] coordinates {(5,4.208)};
    \addplot+[
    boxplot prepared={
        median=3.713,
        upper quartile=4.950,
        lower quartile=2.970,
        upper whisker=5.941,
        lower whisker=2.970
    },
    ] coordinates {(6,3.960)};
    \addplot+[
    boxplot prepared={
        median=3.465,
        upper quartile=4.084,
        lower quartile=2.970,
        upper whisker=4.950,
        lower whisker=2.970
    },
    ] coordinates {(7,3.614)};
    \addplot+[
    boxplot prepared={
        median=3.465,
        upper quartile=3.589,
        lower quartile=2.847,
        upper whisker=3.960,
        lower whisker=2.475
    },
    ] coordinates {(8,3.267)};
    \addplot+[
    boxplot prepared={
        median=2.475,
        upper quartile=3.094,
        lower quartile=1.980,
        upper whisker=3.465,
        lower whisker=1.980
    },
    ] coordinates {(9,2.624)};
    \addplot+[
    boxplot prepared={
        median=2.228,
        upper quartile=2.970,
        lower quartile=1.485,
        upper whisker=3.465,
        lower whisker=1.485
    },
    ] coordinates {(10,2.327)};
  \end{axis}
\end{tikzpicture}
\end{center}
\caption{Vulnerability Distribution per Attempt GPT-4o}
\label{fig:vulnerability_distribution_per_attempt_gpt-4o-2024-08-06}
\end{figure}

Only a subset of the attempts from \autoref{tab:attempt_summary} were executed with GPT-4o.
This was mainly due to the higher costs of using this LLM. We decided to run the most interesting attempts. The essential results are listed in \autoref{tab:results_gpt-4o} and visualized in \autoref{fig:vulnerability_distribution_per_attempt_gpt-4o-2024-08-06}.

All of the tested prompt engineering attempts were successful in reducing the risk of vulnerabilities compared to the baseline. Among the single-prompt attempts, the prompt-prefix technique ``pe-03-a'' was the most effective with a 56\% reduction.

RCI reduced vulnerabilities in the baseline code snippets by 64.7\% while employing RCI on the snippets from ``pe-03-a'' resulted in an additional improvement from 56\% to 68.7\% fewer vulnerabilities compared to the original baseline snippets.

When prompting the model to create vulnerable code in the ``pe-negative'' approach, the ``Scanners Agree Filtered Vulnerable Samples'' metric was 152.7\% above baseline.

The essential results are listed in \autoref{tab:results_gpt-4o}, the box plot in \autoref{fig:vulnerability_distribution_per_attempt_gpt-4o-2024-08-06}.

\begin{table}[htbp]
    \centering
    \renewcommand{\arraystretch}{1.1} 
    \begin{tabular}{|p{3.0cm}|p{1.7cm}|p{1.1cm}|p{1.1cm}|}
    \hline
    \textbf{ID} & \textbf{Filtered \newline Vuln. \newline Sample$^a$ (\%)} & \textbf{diff\textsuperscript{$^b$}(\%)} & \textbf{Vuln. \newline per \newline Sample$^c$} \\
    \hline
rci-from-pe-03-a-iter-1 & 2.33 & +68.7 & 0.43 \\
\hline
rci-from-baseline-iter-1 & 2.62 & +64.7 & 0.40 \\
\hline
pe-03-a & 3.27 & +56.0 & 1.09 \\
\hline
pe-02-a & 3.61 & +51.3 & 1.06 \\
\hline
pe-01-b & 3.96 & +46.7 & 1.05 \\
\hline
pe-01-c & 4.21 & +43.3 & 0.82 \\
\hline
pe-02-b & 4.50 & +39.3 & 1.10 \\
\hline
pe-01-a & 4.80 & +35.3 & 1.07 \\
\hline
baseline & 7.43 & +0.0 & 1.27 \\
\hline
pe-negative & 18.76 & -152.7 & 1.57 \\
\hline
    \multicolumn{4}{p{8cm}}{ \footnotesize
        \vspace{0.1mm} 
        $^a$ Scanners Agree Filtered Vulnerable Samples (average of OFVP)\newline
        $^b$ Relative Difference to Baseline Attempt (higher is better) \newline
        $^c$ Scanners Combined Average Vulnerabilities per Sample (unfiltered)\newline
    }
    \end{tabular}
    \caption{Results for GPT-4o}
    \label{tab:results_gpt-4o}
\end{table}
\vspace{-2mm}

%
%
\section{Discussion}
\label{sec:discussion}

We reflect on the performance of the approaches for each LLM, the importance of sample size, and code functionality.

\begin{figure*}[htbp]
\begin{center}
\begin{adjustbox}{width=\textwidth}
\begin{tikzpicture}
\scriptsize
\begin{axis}[
    height=.2\textheight,
    width=1\textwidth,
    bar width=.18cm,
	ylabel=Vulnerable Percentage,
	enlargelimits=0.05,
    legend style={at={(0.3,1)},
        anchor=north,legend columns=-1},
	ybar = 0pt,
    ymajorgrids = true,
    xtick={1,...,21},
    xticklabels={
        baseline,
        baseline\_100,
        pe-01-a,
        pe-01-b,
        pe-01-c,
        pe-02-a,
        pe-02-b,
        pe-02-c,
        pe-02-d,
        pe-02-e,
        pe-02-f,
        pe-03-a,
        pe-negative,
        ptfscg-cot-iter-1,
        rci-from-baseline-iter-1,
        rci-from-baseline-iter-2,
        rci-from-baseline-iter-3,
        rci-from-pe-03-a-iter-1,
        ptfscg-comprehensive,
        ptfscg-naive-secure,
        ptfscg-persona,
    },
    x tick label style={rotate=30,anchor=north east},
]
\addplot 
	coordinates {
        (1, 5.742574257)
        (2, 5.965346535)
        (3, 6.633663366)
        (4, 7.376237624)
        (5, 7.376237624)
        (6, 6.633663366)
        (7, 6.930693069)
        (8, 7.821782178)
        (9, 7.326732673)
        (10, 6.633663366)
        (11, 7.673267327)
        (12, 6.435643564)
        (13, 10.0990099)
        (14, 6.782178218)
        (15, 4.504950495)
        (16, 4.00990099)
        (17, 3.465346535)
        (18, 4.653465347)
        (19, 5.891089109)
        (20, 6.930693069)
        (21, 5.99009901)
    };
\addplot 
	coordinates {
        (1, 7.326732673)
        (2, 7.653465347)
        (3, 5.742574257)
        (4, 4.603960396)
        (5, 5.04950495)
        (6, 4.603960396)
        (7, 5.198019802)
        (8, 5.940594059)
        (9, 5.495049505)
        (10, 5.445544554)
        (11, 5.841584158)
        (12, 4.059405941)
        (13, 17.42574257)
        (14, 3.96039604)
        (15, 3.861386139)
        (16, 3.168316832)
        (17, 3.168316832)
        (18, 2.97029703)
        (19, 4.158415842)
        (20, 4.158415842)
        (21, 4.95049505)
    };
\addplot+[y filter/.expression={y==0 ? nan : y}]
    coordinates {
        (1, 7.425742574)
        (2, 0)
        (3, 4.801980198)
        (4, 3.96039604)
        (5, 4.207920792)
        (6, 3.613861386)
        (7, 4.504950495)
        (8, 0)
        (9, 0)
        (10, 0)
        (11, 0)
        (12, 3.267326733)
        (13, 18.76237624)
        (14, 0)
        (15, 2.623762376)
        (16, 0)
        (17, 0)
        (18, 2.326732673)
        (19, 0)
        (20, 0)
        (21, 0)
    };
\legend{GPT-3.5-turbo,GPT-4o-mini,GPT-4o}
\end{axis}
\end{tikzpicture}
\end{adjustbox}
\end{center}
\caption{Vulnerability Comparison}
\label{fig:vulnerability_comparison}
\end{figure*}
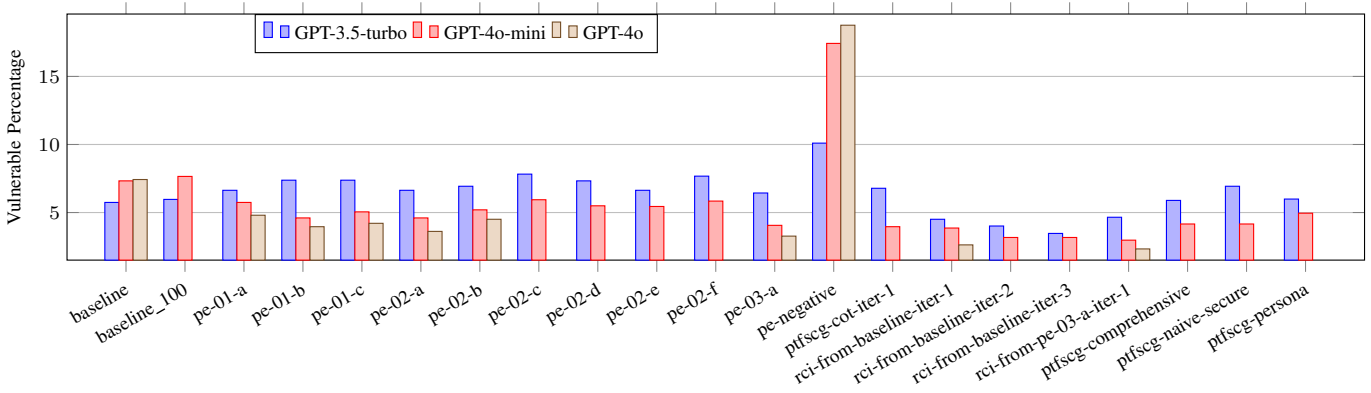

\subsection{Performance Comparison}
\label{subsec:interpretation_of_the_results}

To compare the results of the three LLMs with each other, we illustrated the ``Scanners Agree Filtered Vulnerable Samples'' for each attempt and each LLM in a bar chart in \autoref{fig:vulnerability_comparison}.
This enables the comparison of each LLM's results for every attempt.

The results show an increasing sensitivity to prompt alterations for more advanced models. While generic prompt modifications had no positive impact on GPT-3.5-turbo regarding code security, the more recent GPT-4o-mini model generated significantly more secure code with our prompt additions. We observed the biggest effect on the large GPT-4o model.

The same is true for the model's ability to introduce vulnerabilities on purpose with our ``pe-negative'' adversary attempt where GPT-4o showed the best ability.

Of the simple techniques, which do not require sending additional prompts to the LLM, the prefix ``You are a developer who is very security-aware and avoids weaknesses in the code'' was the most effective in reducing the risk of vulnerable code.

While the RCI technique, which refines the code with additional prompts, was successful for all models, including GPT-3.5-turbo, its effectiveness increased significantly with the more advanced models.

Interestingly, GPT-3.5-turbo had the most secure baseline. The lower average AST height indicates that this might be due to incomplete code. This aligns with the results from Bhatt et al. \cite{Bhatt2023PurpleLC} who found a ``Negative correlation between insecure code test case pass rate and code quality'' and proposed the thesis that ``Models that are more capable at coding tend to be more prone to insecure code suggestions''.

In summary, we can say that RCI provided the best improvement.
It was able to improve the code with additional iterations.
However, this will not work indefinitely; we already observed diminishing improvements by the third iteration on GPT-4o-mini.
It seems very likely that improvements will also stop for other LLMs after a certain number of iterations.
RCI does have the disadvantage of considerably higher costs due to multiple calls with large inputs and outputs.
It is also time-consuming to make multiple calls.
Combining RCI and an attempt like ``pe-03-a'' could boost the results and reduce the number of necessary iterations, reducing cost and time. This approach combines proactive vulnerability prevention in the initial code generation with iterative improvements in code security.

\subsection{Sample Size}
\label{subsec:samle_size}

\autoref{fig:vulnerability_distribution_per_attempt_gpt-3.5-turbo}, \autoref{fig:vulnerability_distribution_per_attempt_gpt-4o-mini} and \autoref{fig:vulnerability_distribution_per_attempt_gpt-4o-2024-08-06} also
highlight the inherent variability in LLM-generated output, which necessitates generating multiple samples for robust evaluation. Code generation can vary significantly across different executions, even with identical prompts on the same model. This variance underscores the importance of sampling to capture a more comprehensive understanding of the model and prompt engineering technique's performance.

In our baseline experiments, we compared the security assessment outcomes from sets of 10 and 100 samples per prompt. While some fluctuations were observed, increasing the sample size to 100 did not result in substantial shifts in the distribution's central tendencies (i.e., average, median, and quartiles) when compared to the results from 10 samples. Given these observations, 10 samples may offer a reasonable balance between computational efficiency and statistical reliability for this specific evaluation. However, it is important to acknowledge that larger sample sizes could reveal additional insights.

\subsection{Functionality}
\label{subsec:functionality}

To assess the impact of our systematic prompt augmentations on the functionality of the generated code, we employed the HumanEval benchmark \cite{Chen2021EvaluatingLL}. 
We converted the prompts from the HumanEval set to our internal dataset format, which allowed us to use our benchmark tool, as described in \autoref{sec:methodology}, to make the prompt augmentations, generate the responses, and extract the code.
To ensure that complete program code would be returned, we added the prefix ``Complete the following code, and output the complete program: \textbackslash n'' to all prompts, prior to the regular prompt augmentation step.
The data file from our benchmark tool, containing the extracted code for all samples, was then converted to the data format expected by the HumanEval framework, and their evaluation scripts were executed to assess the performance of the generated solutions.
Given that this evaluation was not the main focus of our study, we only ran the test for four attempts on GPT-4o-mini, with the usual sample size of 10.
The results are listed in \autoref{tab:humaneval}.

\begin{table}[htbp]
    \renewcommand{\arraystretch}{1.1} 
    \centering
    \begin{tabular}{|l|l|l|}
        \hline
        \textbf{Attempt} & \textbf{pass@1} & \textbf{pass@10} \\
        \hline
        baseline & 86.77\% & 95.12\% \\
        \hline
        pe-03-a & 86.65\% & 94.51\% \\
        \hline
        ptfscg\_rci-from-pe-03-a-iteration-1 & 76.89\% & 93.90\% \\
        \hline
        ptfscg\_rci-from-baseline-iteration-1 & 76.65\% & 94.51\% \\
        \hline
    \end{tabular}
    \newline
    \caption{Functionality test with HumanEval}
    \label{tab:humaneval}
\end{table}

The results of the HumanEval samples show very similar values for pass@10.
The pass@1 values dropped about 10\% for both RCI attempts.
This might be due to the LLM trying to make the code more secure. The default sample size for HumanEval is 200, indicating that 10 samples might not be enough to get accurate numbers. Still, these numbers are a good first indication.
As this was out of scope for our project we did not investigate the matter further.

It is worth mentioning that neither of our scanners detected any vulnerabilities in the generated code for HumanEval.
This might be because the tasks are not designed to provoke any security vulnerabilities and are not very complex, making it less likely to introduce vulnerabilities by mistake.

%
%
\section{Prompt Agent}
\label{sec:prototype_agent}

Our results show significant potential for prompt engineering in secure code generation. The developed ``Prompt Agent'' is an approach to integrating advanced prompt engineering techniques into automated code generation workflows. Its primary goal is to enhance the security of code generated by LLMs such as GPT-4o by systematically improving prompts and incorporating iterative refinement techniques. The prompt engineering techniques were selected based on their performance in our benchmark.

\subsection{Concept}

\begin{figure}[htbp]
\begin{center}
\includegraphics[width=0.95\linewidth]{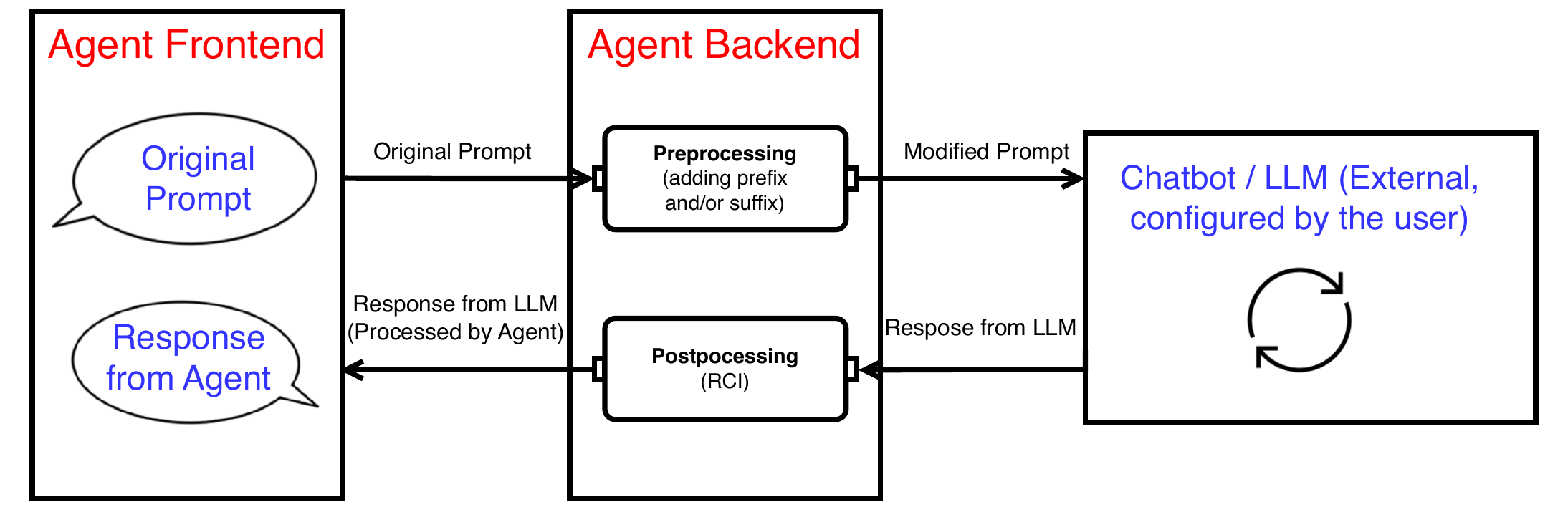}
\end{center}
\caption{Agent Concept}
\label{fig:agent_concept}
\end{figure}

The agent takes a prompt in front end and enhances the prompt in the back end before sending it to the LLM for completion.
After receiving the response, it can optionally apply post-processing using RCI before presenting the output to the user.
Our results indicate that integrating an agent like this into LLM-based tools can significantly reduce the risk of generating vulnerable code.
~\autoref{fig:agent_concept} visualizes the process.

\subsection{Implementation}

Our implementation is based on BetterChatGPT \cite{ztjhzBet72:online}, a popular web front-end for ChatGPT. We have extended the app with two additional features.

\paragraph{Code Security Prefix (Pre\-pro\-cess\-ing the Prompt)}
This feature can be enabled or disabled for each individual prompt.
When enabled, it adds the following prefix to the prompt:
``You are a developer who is very security-aware and avoids weaknesses in the code''.
This was the most successful prefix for GPT-4o-mini and GPT-4o in our evaluation (attempt pe-03-a). As an alternative to manually enabling and disabling the prefix, a classifier could be trained to predict if the user's prompt is likely to generate code, and only augment the prompt in these cases.

\paragraph{Code Security Agent (Postprocessing of the response using RCI)}
The second feature is called ``Code Security Agent'' and can be enabled in the settings. It corresponds to the RCI attempt.
If enabled, the markdown code blocks in the response are intercepted. 

The intercepted blocks are sent to the LLM for security critique and improvement. The code from the improvement response is then extracted and inserted back into the original answer.

If the Code Security Prefix is enabled, the result aligns with the ``ptfscg\_rci-from-pe-03-a-iter-1'' attempt, which was the best in our benchmark for the 4o-family of ChatGPT models. Without the prefix, it aligns with ``ptfscg\_rci-from-baseline-iter-1''. Using the agent with GPT-4o can reduce the number of vulnerabilities on average by 64.7\% without, and 68.7\% with the Code Security Prefix in our scenarios.

\subsection{Usability}

Due to time constraints, we were not able to test and develop our features extensively. While using the agent for some tasks, we noticed that the Code Security Prefix did not noticeably impair the functionality of the chatbots (we used GPT-4o-mini and GPT-4o). 

The Code Security Agent, on the other hand, often delayed the responses by more than 10 seconds, even for a single relatively short code snippet. The reason is that RCI sends additional requests to the LLM. Additionally, we observed RCI to be overly cautious, making the responses too verbose. 
This might be helpful for beginners but can be a deal breaker for advanced programmers.

Based on our first impressions, we think that the Code Security Prefix could be a valuable improvement for programmers of all skill levels using ChatGPT as an assistant. However, post-processing with RCI does not seem to be a viable option for most use cases due to the issues mentioned above. 

More research is required to understand the practical implications of using these techniques for everyday programming tasks. Our agent provides a foundation for further evaluation and is available in our repository.\footnote{\url{https://github.com/mbscit/securecodingprompts}}

\section{Threats to Validity}
\label{sec:threats_to_validity}

\subsection{Construct Validity}
Construct validity concerns whether the study measures what it claims to measure, specifically the efficacy of prompt techniques in enhancing LLM-generated code security.

\textbf{Training Data Influence:}  
Relying on public datasets introduces the risk that models may have been exposed to them during training. If LLMs have encountered these datasets, their performance may reflect prior exposure rather than genuine security capabilities. For instance, GPT-3.5-turbo (trained up to 2021) would not have seen SecurityEval or LLMSecEval datasets but might include source materials such as MITRE CWE and CodeQL documentation. GPT-4o and GPT-4o-mini, trained post-2023, could potentially incorporate these datasets, impacting result interpretation.

\textbf{Scanner Limitations:}  
The study depends on two static scanners to detect vulnerabilities. Discrepancies in scanner performance (e.g., false positives or false negatives) could lead to misinterpretations of security improvements. Additionally, generated code might bypass scanners without addressing underlying vulnerabilities, affecting result reliability.

\textbf{Code Context and Usability:}  
The context in which vulnerabilities are detected may not align with real-world scenarios, as software architecture and operational environments are absent in isolated code snippets. Furthermore, textual warnings in LLM responses were ignored, assuming they might also be overlooked by developers in practice.

\textbf{Impact on Code Functionality:}  
Prompt modifications aimed at enhancing security could inadvertently impair code functionality. This trade-off was only briefly examined, and a deeper exploration is required to understand the balance between functionality and security.

\subsection{Internal Validity}
Internal validity concerns whether the observed outcomes can be attributed to the experimental conditions rather than external or confounding factors.

\textbf{Model and Snapshot Variability:}  
Differences between LLMs, even those from the same vendor, introduce a threat to internal validity. Response variability suggests findings may be model-specific. Additionally, newer snapshots of a model could alter performance, complicating longitudinal comparisons and reproducibility.

\textbf{Prompt Effects:}  
Small variations in prompt design led to significant outcome differences. For instance, the attempts ``ptfscg-persona'' and ``pe-03-a'' reflected similar principles but showed markedly different effectiveness. Combining techniques, such as RCI with ``pe-03-a,'' further demonstrated that systematic tuning can alter results significantly. This variability highlights potential confounding factors introduced by prompt design.

\textbf{Statistical Validation:}
We find notable differences in prompt effectiveness based on observed outcomes and descriptive statistics. However, we have not conducted inferential statistical tests to further validate these results.

\subsection{External Validity}
External validity examines whether findings generalize to broader contexts, such as other LLMs, datasets, or domains.

\textbf{Model-Specific Outcomes:}  
Findings based on OpenAI models (GPT-3.5-turbo, GPT-4o-mini, GPT-4o) may not apply to other LLMs due to significant architectural differences. Techniques effective for one model might fail or even degrade performance in others, limiting cross-model generalizability.

\textbf{Dataset and Language Constraints:}  
The study’s exclusive focus on Python code and a small set of prompts might restrict its applicability to other programming languages or domains. Expanding the dataset and exploring diverse programming tasks would provide broader insights into the generalizability of these techniques.

\textbf{Iterative Generalization:}  
The impact of iterative prompting varies by model. While multiple iterations improved results for some models, others experienced diminishing returns. The optimal number of iterations for different LLMs remains unexplored and could influence the scalability of findings.

\textbf{Scalability of Attempts:}  
This study evaluated a limited subset of prompt variations. Future work should explore a wider range of modifications and combinations, including alternative phrasings and iterative refinements, to identify scalable methods for improving code security.

%
%
\section{Conclusion}
\label{sec:conclusion}

We developed a benchmark to assess how prompt engineering affects code security across GPT models. Our findings show that vulnerability rates and sensitivity to prompt changes vary by model. For GPT-4o-mini and GPT-4o, adding the prefix ``You are a developer who is very security-aware and avoids weaknesses in the code.'' reduced vulnerable code generation by 47\% and 56\%, respectively. In contrast, GPT-3.5-turbo responded negatively to similar security-enhancing prompts. The Recursive Criticism and Improvement (RCI) technique effectively mitigates vulnerabilities across models, with a single iteration fixing 24.5\% (GPT-3.5), 49.5\% (GPT-4o-mini), and 64.7\% (GPT-4o) of flawed snippets. These insights can help end-users and be integrated into LLM-based coding assistants, as demonstrated in our prompt agent concept.

\bibliographystyle{IEEEtran}
\bibliography{IEEEabrv,reference}

\end{document}